\documentclass[%
reprint,
 amsmath,amssymb,
 aps,pra,numerical
]{revtex4-1}
\usepackage{graphicx}
\usepackage{dcolumn}
\usepackage{bm}
\usepackage{multirow}
\usepackage{amsmath}
\usepackage{epstopdf}
\usepackage{natbib}

\begin{document}

\preprint{APS/123-QED}

\title{Experimental investigation of enviroment--induced entanglement \\
using an all--optical setup}

\author{M. H. M. Passos$^1$, W. F. Balthazar$^2$, A. Z. Khoury$^3$, M. Hor--Meyll$^4$, L. Davidovich$^4$, J. A. O. Huguenin$^1$}

\affiliation{%
1- Instituto Ci\^encias Exatas Universidade Federal Fluminense - Volta Redonda - RJ - Brasil\\
2- Instituto Federal do Rio de Janeiro - Volta Redonda - RJ - Brasil\\
3- Institutode F\'{\i}sica - Universidade Federal Fluminense - Niter\'oi - RJ - Brasil\\
4- Instituto F\'{\i}sica - Universidade Federal do Rio de Janeiro - Rio de Janeiro - RJ - Brasil\\
}%


\date{\today}

\begin{abstract}
We investigate the generation of entanglement between two non interacting qubits coupled to 
a common reservoir. An experimental setup was conceived to encode one qubit on the polarization 
of an optical beam and another qubit on its transverse mode. The action of the reservoir is 
implemented as conditional operations on these two qubits, controlled by the longitudinal path 
as an ancillary degree of freedom. An entanglement witness and the two-qubit concurrence are 
easily evaluated from direct intensity measurements showing an excellent agreement with the 
theoretical prediction.
\end{abstract}

\maketitle


\section{\label{introd}Introduction}

Quantum information technologies are challenged by the action of the environment in any kind of physical platform. 
Coherence and entanglement, two essential ingredients for quantum information tasks, are extremely sensitive to noise and 
require efficient strategies for their control and protection. Geometric phase gates \cite{Jones2000,Duan2001}, 
decoherence-free subspaces \cite{Palma1997,Aolita2007} and the nontrivial topology of quantum Hall systems \cite{Kitaev2003}
have been proposed as potential means for implementing noise-robust quantum information protocols. 

The impact of the environment on the dynamics of entanglement has been investigated for a plethora of physical systems beginning with the pioneer work of Rajagopal and Rendell \cite{Rajagopal2001}, where the system under consideration was a pair of coupled quantum dissipative oscillators. Since then several papers have studied the effect of local environments on the dynamics of entanglement of bipartite systems    \cite{Zyczkowski2000,Jakobczyk2003,Dodd2004,Yu2004,Bandyopadhyay2004,Fine2005,Mintert2005,Tolkunov2005,Yu2006,Yu2006a,Santos2006,abliz2006entanglement,Yonac2006,Ficek2006,Yu2007,Ikram2007,Liu2007,Pineda2007,Cunha2007,Marek2008,Gong2008,Al-Qasimi2008,cheon2008bayesian,cormick2008decoherence,Xu2009,Cavalcanti2009,Das2009,Viviescas2010,Salles2008,Barbosa2010,Barreiro2010a} and multipartite systems \cite{Simon2002,Carvalho2004,Hein2005,Lopez2008,Cavalcanti2009,Papp2009,Frowis2011}. Less exploited, but equally important and interesting, is the study of the effects of common environments in bipartite systems \cite{braun2002,kim2002,Schneider2002,Tanas2003,Benatti2006,Ficek2006,Chou2008, Paz2008,Paz2009,Hor-Meyll2009,Auyuanet2010} and multipartite systems \cite{Monz2011}. 

Despite the common sense on the deleterious effects of the environment,  two non-interacting quantum systems may become entangled through the action of a common environment \cite{Hor-Meyll2009}. 
However, the experimental demonstration of this feature is not trivial, since in order to interact with a common environment the systems must be sufficiently close together, implying that the direct interaction between them should be taken into account.  This motivates the search for experimental setups that would allow the demonstration of the onset of entanglement under the sole action of a common environment. Here we present results obtained with an experimental scenario based on linear optics, which circumvents the direct interaction between two qubits, thus allowing the investigation of environment-induced entanglement generation.

Linear optical setups constitute a quite flexible platform for 
testing quantum information principles. They allow encoding of quantum information units (qubits or qudits) in different 
degrees of freedom of the light field, such as polarization, transverse mode or longitudinal path. 
This kind of encoding has already been used to investigate the topological phase acquired by entangled qubit pairs \cite{souza2008quantum,Souza2014}, 
quantum inequalities \cite{Borges2010,Chen2010,Karimi2010,Kagalwala2012,Qian2011,Qian2015,balthazar2016tripartite}, 
quantum cryptography \cite{souza2008quantum,DAmbrosio2012}, quantum image control \cite{caetano2003quantum}, 
quantum gates \cite{Oliveira2005,Souza2010}, quantum simulations \cite{Pinheiro2013}, teleportation schemes 
\cite{chen2009teleportation,Barreiro2010,Khoury2011,Rafsanjani2015,Guzman-Silva2016,Silva2016}, discrete \cite{Barreiro2010} and continuous variables \cite{DosSantos2009,Liu2014} hyperentanglement. 
It has also been used to study entanglement dynamics under the action of the environment \cite{Almeida2007,Salles2008,aiello2007linear}, 
where the role of the environment is usually played by the longitudinal path, while the 
other degrees of freedom represent the quantum systems of interest. Numerous techniques are nowadays available for efficient 
operation of photonic degrees of freedom in quantum protocols
\cite{Ishaaya2008,Nagali2009,KrishnaInavalli2010,Cardano2012,Milione2015,Harris2016,Milione2015,Zhou2015}.

In this work we present an experimental investigation based on the environment-induced entanglement between the polarization and 
the transverse mode of a paraxial beam following the proposal of Ref. \cite{Hor-Meyll2009}. A sequence of conditional operations 
are performed on polarization and transverse mode, controlled by the path degree of freedom that realizes the decoherence 
effects of the environment. An entanglement witness and the concurrence are readily evaluated from intensity measurements 
performed on different output ports of the conditional operations. The experimental results show very good agreement with 
the theoretical predictions. 

The experiment is performed with an intense laser beam, which can be described either as a macroscopic number of photons in a coherent state or simply as a classical electromagnetic field. It captures, nevertheless, the essential features of the phenomenon under investigation here, and is therefore a useful simulation of a single-photon experiment. This kind of simulation provides a test bed for subtle quantum properties, with simple experimental setups, as has been widely discussed in the literature \cite{spreeuw1998classical,souza2007topological,Borges2010,Chen2010,Karimi2010,Kagalwala2012,Qian2011,Qian2015,holleczek2011,aiello2015,ghose2014,toppel2014,mclaren2015,goyal2013,Pinheiro2013,Hor-Meyll2009,spreeuw2001,sun2015,Rafsanjani2015,guzman2016,da2016spin,balthazar2016tripartite}.

The article is organized as follows, in Section \ref{theomod} we review the theoretical 
model for the decoherence process. In Section \ref{exp} we detail the experimental setup and the corresponding procedures. 
The experimental results are presented and discussed in Section \ref{results}. 
Finally, we draw our conclusions in Section \ref{conclusions}.

\section{\label{theomod} Theoretical Model }

\subsection{\label{sec:level2}Entanglement generation}
\label{sec:theoretical model}
In order to illustrate the generation of entanglement by an environment ${\cal E}$, we consider a system ${\cal S}$ composed of two non-interacting qubits in contact with a common bosonic reservoir at zero temperature. Taking $|g\rangle$ as the ground state and $|e\rangle$ as the excited state of each qubit, the collective interaction between the bipartite system and the reservoir is described in the rotating-wave approximation by the Hamiltonian:
\begin{eqnarray}
H= \sum_{i=1}^2 \hbar \omega (|e_{i}\rangle \langle e_{i}|-|g_{i}\rangle \langle g_{i}|) +\sum_{\ell}
\hbar \omega_{\ell}(a^{\dag}_{\ell} a_{\ell}+1/2)\nonumber \\
- i\hbar\sum_{\ell} \sum_{i=1}^{2}
[g_{\ell}(|e_{i}\rangle \langle g_{i}|a_{\ell} - |g_{i}\rangle \langle e_{i}|a^{\dag}_{\ell})].\;\;\;\;\;\;\;\;\;\;\;\;\;\;  
\label{eq1}
\end{eqnarray}
The first and second terms correspond to the free Hamiltonian of the qubits and the reservoir, respectively, and the third term represents the interaction between them, given by means of exchanging excitations. In this expression, $\omega$ stands for the qubit transition frequency, $a^{\dag}_{\ell}$ and $a_{\ell}$ are the creation and annihilation operators associated with mode of the reservoir with frequency $\omega_{\ell}$, and $g_{\ell}$ is the coupling constant. 
We can define the collective operators 
\begin{eqnarray}
\displaystyle S^{+}= \sum_{i=1}^2 |e_{i}\rangle \langle g_{i}|;\;\;\;\;\;\;\displaystyle S^{-}= \sum_{i=1}^2 |g_{i}\rangle \langle e_{i}|
\end{eqnarray}
and describe the evolution of the system $\rho_{\mathcal{S}}(t)$, tracing over the reservoir, by a master equation given by \cite{Gross1982,Ficek2002}:
\begin{eqnarray}
\frac{\partial {\rho}_{\mathcal{S}}(t)}{\partial t}=
\frac{\Gamma}{2}[2S^{-}\rho_{\mathcal{S}}(t)S^{+}
-\rho_{\mathcal{S}}(t)S^{+}
S^{-}-
S^{+}S^{-}\rho_{\mathcal{S}}(t)],\nonumber\\
\label{lindblad}
\end{eqnarray}
where $\Gamma$ corresponds to the spontaneous emission decay rate of a single qubit.

The system evolution given by Eq.~\eqref{lindblad} can be completely reproduced, tracing out the environment, by means of an unitary map which globally describes the time evolution of the system plus the environment. Using both Kraus operators and Choi matrix formalism \cite{kraus1983states,choi1975completely,havel2003robust}, the corresponding map was obtained in \cite{Hor-Meyll2009}:
\begin{eqnarray}
|ee\rangle_{\mathcal{S}}|0\rangle_{\mathcal{E}}&\rightarrow&
A|ee\rangle_{\mathcal{S}}|0\rangle_{\mathcal{E}}+B (|eg\rangle+ |ge\rangle)_{\mathcal{S}}|1_{ee}\rangle_{\mathcal{E}}\nonumber \\
&&C |gg\rangle_{\mathcal{S}}|2\rangle_{\mathcal{E}}\,,\nonumber \\
|eg\rangle_{\mathcal{S}}|0\rangle_{\mathcal{E}}&\rightarrow& D
|eg\rangle_{\mathcal{S}}|0\rangle_{\mathcal{E}}+ E |ge\rangle_{\mathcal{S}}|0\rangle_{\mathcal{E}} + F|gg\rangle_{\mathcal{S}}|1_{eg}\rangle_{\mathcal{E}}\,,\nonumber\\
|ge\rangle_{\mathcal{S}}|0\rangle_{\mathcal{E}}&\rightarrow&
D |ge\rangle_{\mathcal{S}}|0\rangle_{\mathcal{E}}+ E |eg\rangle_{\mathcal{S}}|0\rangle_{\mathcal{E}} + F|gg\rangle_{\mathcal{S}}|1_{eg}\rangle_{\mathcal{E}}\,,
\nonumber\\
|gg\rangle_{\mathcal{S}}|0\rangle_{\mathcal{E}}&\rightarrow&
|gg\rangle_{\mathcal{S}}|0\rangle_{\mathcal{E}}, \label{mapa2}
\end{eqnarray} \\
where $A,B,...,\,F$ are time-dependent coefficients given by:
\begin{eqnarray}
&&A= e^{-\Gamma t},\;\;\;\;\;\; B= \sqrt{\Gamma t e^{-2\Gamma t}},\nonumber\\
&&C=\sqrt{1-e^{-2\Gamma t}-2\Gamma t  e^{-2\Gamma t}},\;\;\;\;\;\; D= \frac{e^{-\Gamma t}+1}{2},\nonumber\\
&&E= \frac{e^{-\Gamma t}-1}{2},\;\;\;\;\;\; F= \sqrt{\frac{1-e^{-2\Gamma t}}{2}}.
\label{timeEqs}
\end{eqnarray}
Detailed calculations of the coefficients can be found in \cite{Hor-Meyll2009}.
The physical interpretation of the above unitary map is straightforward. For example, consider the first line, where both qubits are initially excited, represented by the state $|ee\rangle_{\cal{S}}$ and the environment is in the vaccum state $|0\rangle_{\cal{E}}$. According to the map, this state evolves to a coherent superposition of three states, corresponding to the three different possible physical situations. The state $|ee\rangle_{\cal{S}}|0\rangle_{\cal{E}}$, associated to probability amplitude  $A$, indicates that system and environment do not exchange excitations;  the state  $(|eg\rangle+ |ge\rangle)_{\mathcal{S}}|1_{ee}\rangle_{\mathcal{E}}$, associated with probability amplitude $B$, implies that one of the qubits decays emitting one excitation into the environment; and, finally, the state $|gg\rangle_{\mathcal{S}}|2\rangle_{\mathcal{E}}$, associated with probability amplitude $C$, corresponds to the decay of both qubits emitting two excitations into the environment. In the second (or third line), where one of the qubits is initially excited while the other one is in the ground state, there are three possible situations. There is a probability amplitude $D$ of both remaining in the same state, a probabilty amplitude $E$ that the qubits exchange the excitation, which is mediated by the common environment, and finally a probability amplitude $F$ that the excited qubit decays emitting the excitation into the environment. The interpretation of the last line is trivial, since both qubits start in the ground state, i.e., no evolution is possible. From our experimental perspective, it is more convenient to represent the above dynamics using the formalism of unitary maps, since they can easily  be implemented by an optical apparatus, as will be shown in the following.  The situation we are interested in, entanglement generation by the environment, is depicted in the second (or third) line of the map.  There, an initially separable state $|eg\rangle_{\mathcal{S}}$ (or $|ge\rangle_{\mathcal{S}}$) evolves towards an entangled two-qubit state, through the interaction with the common environment.

\subsection{\label{sec:level2}Entanglement witness}
\label{sec:entanglement measurement}
In order to quantify the amount of entanglement that is induced by the reservoir in the two-qubit system $\rho_{\cal{S}}(t)$, we use the  \textit{concurrence} \cite{wootters1998entanglement}, given by:
\begin{equation}
{\cal C} = {\rm max}\{0,\Lambda\},
\label{concurrence}
\end{equation}
where $\Lambda=\sqrt{\lambda_{1}}-\sqrt{\lambda_{2}}-\sqrt{\lambda_{3}}-\sqrt{\lambda_{4}}$ and $\lambda_{i}$ are the eigenvalues, in decreasing order, of the hermitian matrix:
\begin{equation}
\rho_{\cal{S}}(t)(\sigma_{y}\otimes\sigma_{y})\rho_{\cal{S}}(t)^*(\sigma_{y}\otimes\sigma_{y}),
\label{eq6}
\end{equation}
where $\sigma_{y}$
is the Pauli matrix and the conjugation is realized in the
$\{|ee\rangle, |eg\rangle, |ge\rangle, |gg\rangle\}$ basis. For a separable state ${\cal C}=0$ and for a maximally entangled state  ${\cal C}=1$.

We investigate environment-induced entanglement generation in the physical situation where the initial state of the system is the separable state $\rho_{\cal{S}}(0) = |eg\rangle \langle eg|$, and whose evolution is described by the second line of the unitary map in Eq.~(\ref{mapa2}). After evolution during a time $t$, the state of the system $\rho_{\mathcal{S}}(t)$, upon tracing out the environment states,  is given by: 
\begin{eqnarray}
\rho_{\mathcal{S}}(t)=&&\left(\frac{e^{-\Gamma t}+1}{2}\right)^{2}|eg\rangle\langle eg|+\left(\frac{e^{-\Gamma t}-1}{2}\right)^{2}|ge\rangle\langle ge|\nonumber\\
&&+\frac{e^{-2\Gamma t}-1}{4}(|eg\rangle\langle ge| +|ge\rangle\langle eg|) \nonumber\\
&&+\frac{1-e^{-2\Gamma t}}{2}|gg\rangle\langle gg|. \label{rhodete}
\end{eqnarray}
One can easily check that the initial separable state evolves to an entangled state for any time $t>0$. Indeed, a straightforward calculation gives:
\begin{eqnarray}
\Lambda=\frac{1}{2}(1-e^{-2\Gamma t}).\label{eq8}
\end{eqnarray}

Since there is no direct interaction between the qubits, this entanglement is due solely to the common interaction with the environment. So, in order to investigate this effect, it suffices to implement either the second or third line of the unitary map in Eq.~\eqref{mapa2}.

To understand qualitatively the emergence of entanglement under the influence of the environment, it is convenient to express the initial state $|eg\rangle$ as a superposition of the  maximally entangled states $|\psi^\pm\rangle$: 
\begin{eqnarray}
|eg\rangle=\frac{1}{\sqrt{2}}(|\psi^-\rangle +|\psi^+\rangle),
\label{eg}
\end{eqnarray} 
where
\begin{eqnarray}
|\psi^{\pm}\rangle=\frac{1}{\sqrt{2}}(|eg\rangle \pm |ge\rangle).\nonumber\\
\label{bell_psi}
\end{eqnarray} 
Now, according to Eq.~(\ref{lindblad}),  $|\psi^{-}\rangle$ does not evolve -- it is a decoherence-free state -- while $|\psi^{+}\rangle$ decays asymptoptically to the ground state $|gg\rangle$. Therefore, in the limit $t\rightarrow \infty$ the system approaches the state:
\begin{equation}
\rho_{est}=
\frac{1}{2}|\psi^-\rangle\langle \psi^-|+\frac{1}{2}|gg\rangle\langle gg|.
\label{rhoest}
\end{equation}
In fact, concurrence of this state is equal to $1/2$, in accordance with the expected value for $t\rightarrow \infty$ in Eq.~(\ref{eq8}).  Entanglement arises from the  maximally entangled state $|\psi^-\rangle$.

In general, determination of concurrence requires full tomographic reconstruction of the physical states, which is extremely time consuming. However, in some situations in which one has some {\it a priori} knowledge of the state space involved, it is possible to work with  an {\it entanglement witness}, an observable whose mean value allows for signaling the presence of entanglement. Let $W$ be an entanglement witness. Then, for any separable state $\rho_{sep}$ we must have ${\rm Tr}(W\rho_{sep})\geq0$ and at least for some set of entangled states $\rho_{ent}$, we will have 	${\rm Tr}(W\rho_{ent}) < 0$ and the states belonging to this set are said to be detected by the witness. An optimal witness is the one which detects all entangled states in the space state considered in the problem. 

Fortunately, there is an optimal entanglement witness for the family of states considered in Eq.~({\ref{rhodete}), which was obtained in Ref.~\cite{Hor-Meyll2009}, represented here in its eigenvectors basis $\{|\psi^+\rangle, |ee\rangle, |\psi^-\rangle, |gg\rangle\}$:
	\begin{eqnarray}
	W&=&\left(1+\frac{1}{\sqrt{2}}\right)P_{ee}+\frac{1}{\sqrt{2}}P_{\psi^+}-\nonumber\\
	&&\frac{1}{\sqrt{2}}P_{\psi^-}+\left(1-\frac{1}{\sqrt{2}}\right)P_{gg},
	\label{diagonalwitness}
	\end{eqnarray}
	where $P_{\phi}\equiv |\phi\rangle \langle \phi|$ is the projector on state $|\phi\rangle$.
	This observable surpasses the role of a witness, {\it i.e.},  simply {\it detection} of entanglement. Indeed, its mean value is proportional to $\Lambda$, so  it {\it quantifies} the entanglement all over the time. The mean value of $W$ is related to $\Lambda$, and consequently to the concurrence,  by the formula:
	\begin{equation}
	\Lambda=\frac{{\rm Tr}[W\rho_{\mathcal{S}}(t)]}{(1-\sqrt{2})}\,. \label{Ccalc}
	\end{equation}

	Therefore, to experimentally quantify the amount of entanglement induced by the environment for any instant of time, we just need to measure a single observable avoiding full tomographic reconstruction of the state.

	\section{\label{exp}Experimental Setup}
	
	In order to  experimentally investigate the environment-induced entanglement, we implement an all-optical setup  based on the proposal of Ref. \cite{Hor-Meyll2009}. There, the proposed implementation of the unitary quantum map, corresponding to the evolution of the two-qubit system interacting with the common environment, exploits different degrees of freedom of a single photon. In our experimental realization, the map is implemented by encoding qubits and environment in optical modes of a laser beam. The first qubit is encoded in the polarization degree of freedom. The horizontal polarization ($H$) represents the ground state ($\left|H \right> \equiv \left|g \right>$) and the vertical polarization ($V$) the  excited one ($\left| V \right> \equiv \left| e \right>$). The second qubit is encoded in the first order transverse mode degree of freedom. We use first-order Hermitian-Gaussian modes. The ground state is encoded in the $HG_{01}$ mode. We define $\left|g \right> \equiv \left| HG_{01} \right> \equiv \left|h \right>$, where $h$ stands for horizontal nodal line. Finally, the excited state of second qubit is encoded  in the mode  $HG_{10}$ and we have the similar relation $\left|e \right> \equiv \left| HG_{10} \right> \equiv \left|v \right> $. Here, $v$ stands for vertical nodal line. The correspondence between two-qubit states and optical-modes is summarized below:
	\begin{eqnarray}
	\left| ee \right>  &\equiv& \left|Vv \right>, \nonumber\\
	\left| eg \right> &\equiv& \left|Vh \right>, \label{equivmode}\\
	\left| ge \right> &\equiv& \left|Hv \right>, \nonumber\\
	\left| gg \right>  &\equiv& \left|Hh \right>. \nonumber
	\end{eqnarray}
The evolution we are interested in [second line of Eq.~\eqref{mapa2}] can be written in terms of the optical modes as
	
	\begin{eqnarray}
	\label{rmap}
	|Vh\rangle_{\mathcal{S}}|0\rangle_{\mathcal{E}} \rightarrow  D
	|Vh\rangle_{\mathcal{S}}|0\rangle_{\mathcal{E}}+ E |Hv\rangle_{\mathcal{S}}|0\rangle_{\mathcal{E}} + F|Hh\rangle_{\mathcal{S}}|1_{Vh}\rangle_{\mathcal{E}}.\nonumber\\
	\end{eqnarray}
The different states of the environment are encoded in different paths of the laser beam, to be defined in the description that follows.

	The experimental setup is sketched in Fig.~\ref{Circuit}. A diode pumped solid state (DPSS) laser beam ($532~nm$, $1.5~mW$, vertically polarized) is incident on a S-wave plate ($SP$), producing the so-called vector vortex beam $\frac{1}{\sqrt{2}} (\left|Vh \right> + \left|Hv\right>)$. After being transmitted through $PBS1$, state $\left|Hv\right>_{\mathcal{S}}$ is blocked, while state $\left|Vh\right>_{\mathcal{S}}$, corresponding to input state $\left|eg\right>$, is reflected and incident on a spatial filter ($SF$) in order to improve the fidelity of the transverse spatial mode. 	We define the path following the spatial filter as the vacuum state of the environment, so we end up with the initial state given by ${\left|Vh\right>}_{\mathcal{S}}\left|0\right>_{\mathcal{E}}$. After that,  a half wave plate ($HWP1$), aligned at an angle $\theta_1$ with respect to the vertical polarization, performs the following transformation:
	\begin{eqnarray}
	\left|Vh\right>_{S}\left|0\right>_{\mathcal{E}} \rightarrow \left[ \cos(2\theta_1) \left|Vh\right> + \sin(2\theta_1)\left|Hh\right> \right]_{S}\left|0\right>_{\mathcal{E}}.\nonumber\\
	\end{eqnarray}
	The ${\left|Vh\right>}_{\mathcal{S}}$ component  is reflected through $PBS2$ following the auxiliar path ${\left|0^\prime\right>}_\mathcal{E}$, shown in Fig.~\ref{Circuit}, where a set of three mirrors is used such that this component is incident on $PBS3$. One of these mirrors is placed on a translation stage with a micrometer vernier in order to perform a fine adjustment of the propagation path. In this way, we can introduce a dynamical phase $\Delta \phi$ in the propagation of path ${\left|0^\prime\right>}_\mathcal{E}$. The $\left|Hh\right>_{\mathcal{S}}$ component is transmitted through $PBS2$ following the path $\left|1\right>_\mathcal{E}$  and is incident on a Dove prism $DP1@\theta_2$ aligned at an angle $\theta_2$ with respect to the vertical orientation. After $DP1@\theta_2$, the state of the system plus environment in path $\left|1\right>_\mathcal{E}$ is described by 
	
	\begin{equation}
	\left|Hh\right>_{S}\left|1\right>_{\mathcal{E}} \rightarrow \left[ \cos(2\theta_2) \left|Hh\right> + \sin(2\theta_2)\left|Hv\right> \right]_{S}\left|1\right>_{\mathcal{E}}.
	\end{equation}
	
	After $DP1@\theta_2$, a Mach-Zehnder interferometer with an additional mirror (MZIM)  performs a mode parity selection \cite{Sasada2003}. The MZIM is composed by two 50/50 beam splitters ($BS$) and three mirrors. In the upper arm, two mirrors perform the double reflection required for parity selection. On the lower arm, the mirror is mounted over a piezoelectric ceramic ($PZT$), in order to control the phase difference  between the two arms.  
	
	By adjusting the phase difference between the MZIM arms, we can separate the even component $\left|Hh\right>_{\mathcal{S}}$ (following path $\left|1\right>_\mathcal{E}$)  from the odd  component $\left|Hv\right>_{\mathcal{S}}$ (following path $\left|0^{\prime\prime}\right>_\mathcal{E}$). So, after the MZIM, we end up with the following state:   
	
	\begin{eqnarray}
	\left|Hh\right>_{S}\left|1\right>_{\mathcal{E}} \rightarrow \cos(2\theta_2) \left|Hh\right>_{S}\left|1\right>_{\mathcal{E}} + \sin(2\theta_2)\left|Hv\right>_{S} \left|0^{\prime\prime}\right>_{\mathcal{E}},\nonumber\\
	\end{eqnarray}
where paths $\left|1\right>_{\mathcal{E}}$ and $\left|0^{\prime\prime}\right>_{\mathcal{E}}$ are shown in Fig.~\ref{Circuit}.
In $PBS3$, the path components $\left|0^{\prime\prime}\right>_\mathcal{E}$ and $\left|0^{\prime}\right>_\mathcal{E}$ are coherently combined and identified as vacuum state of the environment ${\left|0\right>}_\mathcal{E}$, as described by the unitary map of Eq.~\eqref{mapa2}.

	Finally, the entire transformation performed by the Preparation circuit, shown in Fig.~\ref{Circuit}, can be written as 
	
	\begin{equation}
	\begin{split}
	\left|Vh\right>_{S}\left|0\right>_{\mathcal{E}} \rightarrow [\cos(2\theta_1)\left|Vh \right> + \sin(2\theta_1)\sin(2\theta_2)\left|Hv \right>]_{S}\left|0 \right>_\mathcal{E} \\
	+ \sin(2\theta_1)\cos(2\theta_2)\left|Hh \right>_{S}\left|1 \right>_\mathcal{E}.
	\end{split}
	\label{generalst}
	\end{equation}
	Identifying  $D=\cos(2\theta_1)$, $E=\sin(2\theta_1)\sin(2\theta_2)$, and $F=\sin(2\theta_1)\cos(2\theta_2)$ we are able to simulate the second line of Map \eqref{mapa2}, as desired.	
	\begin{figure}[!htp]
		\centering
		\includegraphics[scale=6.9,trim=0.21cm 0cm 0cm 0cm, clip=true]{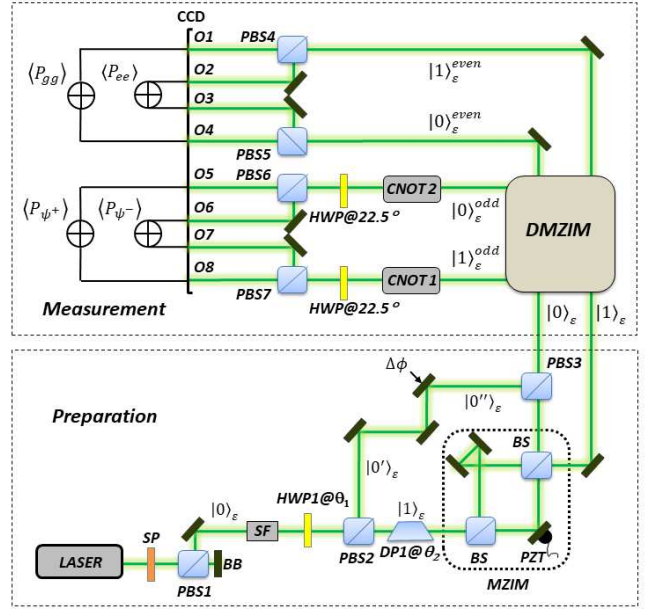}
		\caption{Experimental setup.}
		\label{Circuit}
	\end{figure} 
Note that the evolution time will be set by adjusting $\theta_1$ and $\theta_2$ accordingly.  In order to respect the time evolution for the coefficients $D$, $E$, and $F$, taking into account the parity of sine and cosine functions, the angle $\theta_1$ ($\theta_2$) must vary in counterclockwise (clockwise) direction.

	The beams propagating in path $\left| 0 \right>_\mathcal{E}$ and $\left|1 \right>_\mathcal{E}$ are then directed to the measurement circuit. They are sent to a double input/output MZIM (DMZIM), sketched in Fig.~\ref{domzim}. The DMZIM is composed by a large input $50/50$ beam splitter ($BS1$), two regular output beam splitters ($BS2$ and $BS3$) and three mirrors. The phase difference between the interferometer arms will be adjusted by an unique mirror with a $PZT$. 
	 
	The two components resulting from splitting of path $\left| 0 \right>_\mathcal{E}$ in $BS1$ are coherently recombined in the $BS2$ (solid line). The same for path $\left|1 \right>_\mathcal{E}$ regarding  $BS3$ (dashed line).  Then, with proper adjustment of the phase difference between arms, the even and odd components of the path $\left| 0 \right>_\mathcal{E}$and $\left| 1 \right>_\mathcal{E}$ will be separated by the DMZIM and, as shown in Fig.~\ref{domzim}, redefined as $\left| 0 \right>_\mathcal{E}^{even}$, $\left| 0 \right>_\mathcal{E}^{odd}$, $\left| 1 \right>_\mathcal{E}^{even}$ and  $\left|1 \right>_\mathcal{E}^{odd}$,  respectively.  This apparatus is important because it simultaneously analyzes two incoming beams in their even and odd components, separating them into two independent outputs for the same phase difference between the arms of the DMZIM.  

\begin{figure}[!htp]
\centering
\includegraphics[scale=0.7]{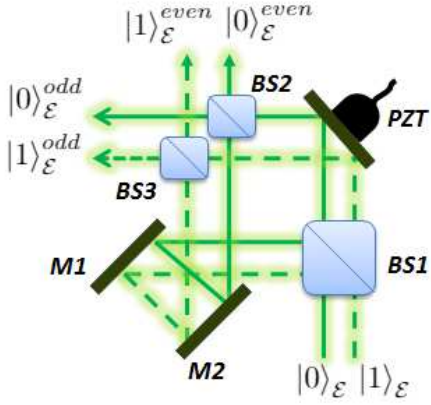}
\caption{Double input/output MZIM.}
\label{domzim}
\end{figure}

Following the scheme proposed in Ref.~\cite{Hor-Meyll2009}, the components $\left|1 \right>_\mathcal{E}^{even}$ and $\left|0 \right>_\mathcal{E}^{even}$ are directed to $PBS4$ and $PBS5$, respectively,  as shown in the Measurement circuit of Fig.~\ref{Circuit}.  As $PBS$ transmits $H$-polarized beams and reflect $V$-polarized ones, component  $\left|Hh \right>_{S}$ of the path  $\left| 1 \right>_\mathcal{E}^{even}$ will be detected in output $O1$ and component $\left|Vv \right>_{S}$ in $O2$. Note that the component $\left|Vv \right>_{S}$ is not present in the state given by the second line of Eq.~\eqref{mapa2}, so it is expected zero intensity in  output $O2$. Similarly, for path $\left| 0 \right>_\mathcal{E}^{even}$, measurements will be performed in outputs $O3$ and $O4$. Again, as path $\left| 0 \right>_\mathcal{E}$ has only odd components, no intensity is expected from outputs of $PBS5$. The components $\left|1 \right>_\mathcal{E}^{odd}$ and $\left|0 \right>_\mathcal{E}^{odd}$ are directed to two Controlled-Not gates, $CNOT1$ and $CNOT2$, respectively, as shown in Fig.~\ref{Circuit}. 

\begin{figure}[!htp]
\centering
\includegraphics[scale=0.65,trim=8.5cm 9cm 10cm 5cm, clip=true]{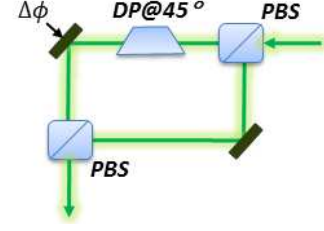}
\caption{CNOT circuit}
\label{cnot}
\end{figure}

The interferometer implementing the $CNOT$ gate is shown in Fig.~\ref{cnot}. The polarization degree of freedom is used as the control-bit and the transverse mode degree of freedom as the target-bit. A $H$-polarized beam is  transmitted through the first $PBS$ and its transverse mode undergoes a $90^\circ$ rotation by means of a Dove prism aligned at $45^\circ$ with respect to the vertical orientation ($DP@45^\circ$). A mirror mounted in a translation stage redirects the $H$-polarized beam to the second $PBS$ and allows the control of the phase difference $\Delta \phi$ between the two arms. The $V$-polarized beam is reflected through the first $PBS$ and by the mirror and simply recombined with the $H$-polarized beam in the second $PBS$. 

After each $CNOT$ gate, a half wave plate ($HWP@22.5^\circ$), aligned at an angle $22.5^\circ$ with respect to the vertical polarization acts as a Hadamard gate. Finally, $PBS6$ ($PBS7$) in path $\left| 0 \right>_\mathcal{E}$ ($\left| 1 \right>_\mathcal{E}$) implies that the intensity of component $\left|Hv \right>_{S}$ will be measured in output $O5$ ($O8$) and that of component $\left|Vh \right>_{S}$ will be measured in output $O6$ ($O7$). The combination of $CNOT$, $HWP@22.5^\circ$ and $PBS$ indeed amounts to the following transformations:
\begin{eqnarray}
|\psi^+\rangle_{S} \rightarrow \left|Hv \right>_{S}\nonumber\\
|\psi^-\rangle_{S} \rightarrow \left|Vh \right>_{S}
\label{tomography}
\end{eqnarray}
so, outputs $O5$ plus $O8$ ($O6$ plus $O7$) measures the intensity corresponding to state $|\psi^+\rangle$ ($|\psi^-\rangle$) outgoing the Preparation circuit.

All measurement outputs are directed to a screen and a Charge Coupled Device (CCD) camera is used to register simultaneously the image of all outputs in an unique frame. The intensity $I_{j}$ associated with each output $Oj ~ (j = 1,...,8)$ is obtained by integrating the respective images gray scale distribution.    

The population of each state $\left< P_\phi\right>$, where $\phi = |ee\rangle, |gg\rangle, |\psi^+\rangle, |\psi^-\rangle$, is obtained by adding two normalized outputs intensities $I_j/I_T$, where $I_T = \sum_{j=1}^8 I_j$, according to:
\begin{eqnarray}
\left< P_{gg} \right> = \frac{I_1 + I_4}{I_T}, \nonumber \\
\left< P_{ee} \right> = \frac{I_2 + I_3}{I_T}, \label{pop}\\
\left< P_{\psi^+} \right> = \frac{I_5 + I_8}{I_T},\nonumber\\
\left< P_{\psi^-} \right> = \frac{I_6 + I_7}{I_T} \nonumber.
\end{eqnarray}
This procedure is equivalent to the measurement of two beams in an unique detector, as proposed in Ref.~\cite{Hor-Meyll2009}. 

In order to obtain $\Lambda$, we calculate the trace of $W\rho_{\mathcal{S}}(t)$ which, according to Eq.~\eqref{diagonalwitness}, will be given by 
\begin{eqnarray}
Tr[W\rho_{\mathcal{S}}(t)] =\left(1+\frac{1}{\sqrt{2}}\right)\left< P_{ee}\right>+\frac{1}{\sqrt{2}}\left< P_{\psi^+}\right>-\nonumber\\
\frac{1}{\sqrt{2}}\left< P_{\psi^-}\right>+\left(1-\frac{1}{\sqrt{2}}\right)\left< P_{gg}\right>.
\label{trace}
\end{eqnarray}
By using Eq.~\eqref{trace},  $\Lambda$ is calculated from Eq.~\eqref{Ccalc}.

\section{\label{results}Results}

Let us first consider the case when $t=0$, which means that the system is still unaffected by the environment. It corresponds to angles $\theta_1=\theta_2 =0^\circ$.  In this situation, the beam is reflected in $PBS2$ and $PBS3$, and the initial state $\left|Vh\right>_{S}\left|0\right>$ emerges unaltered from the Preparation circuit. After the parity selection performed by the DMZIM, this odd mode component passes unaltered through  $CNOT2$. Finally, by the action of the $HWP@22.5^\circ$ it is equally split by $PBS6$ resulting in the same intensity for outputs $O5$ and $O6$. The resulting images are presented in Figure~\ref{images}-a. Indeed, only outputs $O5$ and $O6$ are illuminated while
all other outputs are not. We observe a very low noise corresponding to the even output $O4$ of DMZIM due its imperfect visibility, which we measure to be $93\%$. It is worth to mention that the preparation MZIM has a visibility around $97\%$. For the DMZIM, the alignment is slightly more delicate due the unique input port.    
\begin{figure}[!htp]
\centering
\includegraphics[scale=0.76,trim=5.4cm 0cm 5cm 0cm, clip=true]{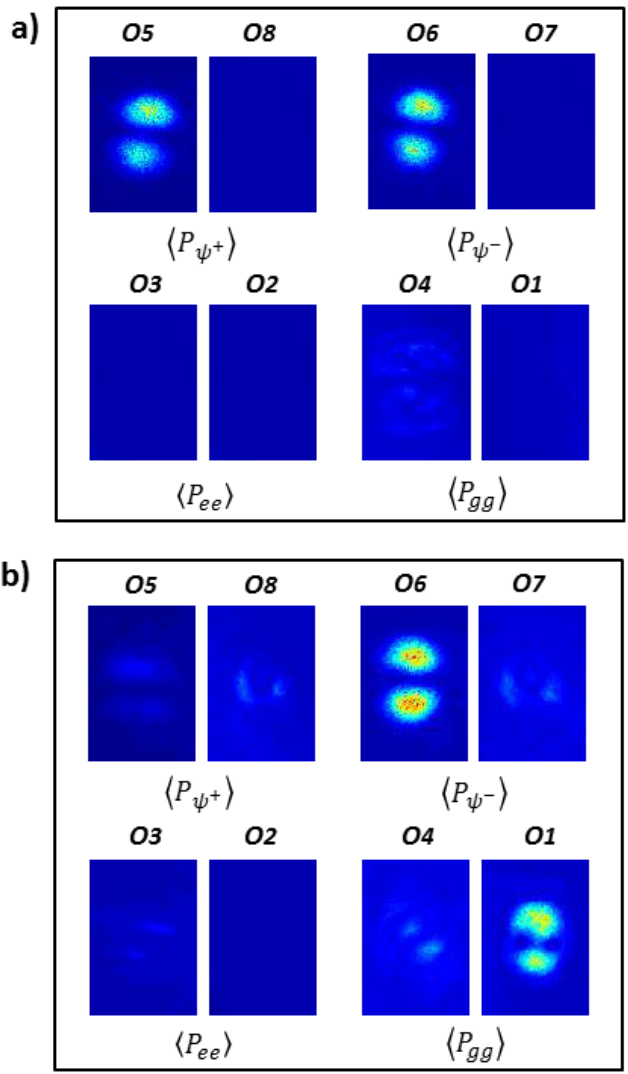}
\caption{Output images for a) initial state ($t=0$) and b) asymptotic state ($t\rightarrow\infty$). The population, shown at the bottom of each pair of images, is obtained by summing the grey scale distribuition of the corresponding outputs.}
\label{images}
\end{figure}
Table \ref{table1}-a presents the expected (Theo) and measured (Exp) population for each state and the resulting $\Lambda$ for the case $t=0$. The experimental populations were obtained from the intensities shown in Fig.~\ref{images}-a by using Eq.~\eqref{pop}.
$\Lambda$ was calculated by means of Eq.~\eqref{Ccalc} resulting in $\Lambda = -0.02 \pm 0.02$. Despite the negative value for $\Lambda$, we match, within the error bar, the theoretically expected value $(\Lambda=0)$. This negative value can be understood as a consequence of experimental errors. In addition, it is worth to mention that the values of the populations are also in very good agreement with the theoretical prediction for the initial separable state $|eg\rangle_{S}$. The error bar of $\Lambda$ was obtained by error propagation in $Tr[W\rho_{\mathcal{S}}]$ [Eq.\eqref{Ccalc})] taking into account the errors in population values due to intensity uncertainties given by the sensitivity of the CCD camera ($\pm2\%$).

\begin{table}[!htp]
\centering
\caption{Theoretical (Theo) and Experimental (Exp) populations and the resulting value for  $\Lambda$ for two pairs of angles. The errors of the populations ($\pm2\%$) and $\Lambda$ ($\pm 0.02$) are omitted.}
\label{table1}
\begin{tabular}{lcc|c|c|c|c|c|}
	\cline{4-8}                                                   &                                                                                                                      &                & \textbf{$\left< P_{\psi^+} \right>$} & \textbf{$\left< P_{\psi^-} \right>$} & \textbf{$\left< P_{ee} \right>$} & \textbf{$\left< P_{gg} \right>$} & \textbf{$\Lambda$} \\ \hline
	\multicolumn{1}{|l|}{\multirow{2}{*}{\textbf{a)}}} & \multicolumn{1}{c|}{\multirow{2}{*}{\textbf{\begin{tabular}[c]{@{}c@{}}$\theta_1 = 0^\circ $ \\ $\theta_2 = 0^\circ $\end{tabular}}}} & \textbf{Theo} & $0.50$         & $0.50$         & $0.00$         & $0.00$         & $0.00$       \\ \cline{3-8} 
	\multicolumn{1}{|l|}{}                             & \multicolumn{1}{c|}{}                                                                                                & \textbf{Exp}  & $0.46$         & $0.44$         & $0.00$         & $0.05$         & $-0.02$       \\ \hline
	\multicolumn{1}{|l|}{\multirow{2}{*}{\textbf{b)}}} & \multicolumn{1}{c|}{\multirow{2}{*}{\textbf{\begin{tabular}[c]{@{}c@{}}$\theta_1 = 30^\circ $\\ $\theta_2 = -18^\circ $\end{tabular}}}} & \textbf{Theo} & $0.00$         & $0.50$         & $0.00$         & $0.50$         & $0.50$       \\ \cline{3-8} 
	\multicolumn{1}{|l|}{}                             & \multicolumn{1}{c|}{}                                                                                                & \textbf{Exp} & $0.01$         & $0.57$         & $0.05$         & $0.37$         & $0.49$       \\ \hline
\end{tabular}
\end{table}

Another interesting case is the behavior for $t \rightarrow \infty$, when the interaction with the environment is maximal. In this situation
we have $\theta_1=30^\circ$ and $\theta_2= -18^\circ$. Due to $HWP@30^\circ$ and $PBS2$ the  system state acquires a $\left| Hh \right>_{\mathcal{S}}$ component, which is directed to the path labelled $\left| 1 \right>_\mathcal{E}$. After passing through the $DP1@\theta2$, another component, namely $\left| Hv \right>_{\mathcal{S}}$, is added to the general state. The $MZIM$ selects the even mode component $\left| Hh \right>_{\mathcal{S}}$ to exit in path $\left| 1 \right>_\mathcal{E}$ and the odd component $\left| Hv \right>_{\mathcal{S}} $ to exit in path $\left| 0^{\prime\prime} \right>_\mathcal{E}$. This last component is coherently superposed with the component $\left| Vh \right>_{\mathcal{S}}$ coming from path  $\left| 0^\prime \right>_\mathcal{E}$ in $PBS3$, resulting in the path labelled $\left| 0\right>_\mathcal{E}$. 
The control of the phase difference $\Delta\phi$ between these paths is critical for achieving the proper coherent superposition. In the meaurement circuit, DMZIM splits even and odd components of paths  $\left| 0 \right>_\mathcal{E}$ and  $\left| 1 \right>_\mathcal{E}$. The even component $\left| Hh \right>_{\mathcal{S}}$ of path $\left| 1 \right>_\mathcal{E}$ is transmitted in $PBS4$ so that output $O1$ is illuminated. After passing through the $CNOT2$, $HWP@22.5^\circ$ and $PBS6$, the state $|\psi^{-}\rangle_{\mathcal{S}}$, resulting from the coherent superposition of the odd components $|Hv\rangle_{\mathcal{S}}$ and $|Vh\rangle_{\mathcal{S}}$ on path $|0\rangle_{\mathcal{E}}$, is converted in the state $\left| Vh \right>_{\mathcal{S}} $ (Eq.~\ref{tomography}), so that output $O6$ is illuminated. Indeed, as shown in Fig.~\ref{images}-b, only outputs $O1$  and $O6$ have no negligible intensities. All other outputs present residual intensities due to imperfections of the optical components and the limited visibility in the DMZIN, which in this case is around $90\%$.
Table \ref{table1}-b shows the populations experimentally calculated from gray scale intensities of Fig.~\ref{images}-b and the respective theoretical predictions. Again, we observe a very good agreement between the measured and expected populations, leading us to $\Lambda=0.49 \pm 0.02$  extremely close to the theoretical value ($\Lambda=0.50$). This case corresponds to the maximum entanglement that can be induced in the system by the interaction with the environment.

In order to study the general evolution behavior of entanglement induced by the environment, we have taken eleven pairs of angles $\theta_1$ and $\theta_2$ to obtain $\Lambda$ for eleven different instants of time. For this purpose, it is convenient to define the time-dependent parameter $p=1-e^{-\Gamma t}$. For $t=0$, $p=0$, corresponding to situation depicted in Table \ref{table1}-a. For the case of maximum entanglement ($t\rightarrow \infty$), $p=1$, corresponding to the results in Table \ref{table1}-b. We have set the values of $\theta_1$ and $\theta_2$ resulting in eleven regularly spaced values of $p$ between $0$ and $1$. In Fig.~\ref{concurrenceresults} we plot $\Lambda$ inferred from the measurement results (triangles) and the corresponding theoretical prediction for the concurrence (solid line) as a function of $p$. The agreement between them is remarkable over the whole interval $0\leq p\leq1$.        

\begin{figure}[h!]
\centering
\includegraphics[scale=0.39,trim=0cm 0cm 0cm 0cm, clip=true]{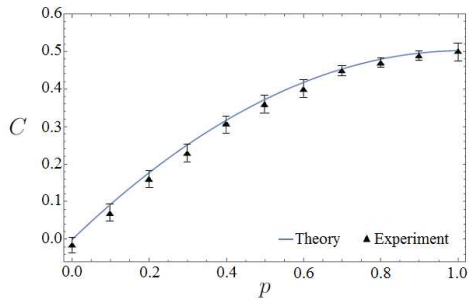}
\caption{Concurrence $C=\max\{0,\Lambda\}$ as a function of the parameter $p=1-e^{-\Gamma t}$. Experimental values for $\Lambda$  are represented by triangles,  while the solid line  represents the theoretical prediction for $C$.}
\label{concurrenceresults}
\end{figure}

\section{\label{conclusions}Conclusions}

We have experimentally investigated the induction of entanglement by a common environment acting on a two-qubit system, using the structural non-separability between the transverse modes and the polarization of a classical laser beam. The optical setup allows the investigation of the sole effect of the environment, since there is no direct interaction between the qubits. Furthermore,  an optimal entanglement witness has allowed the quantification of entanglement over the entire evolution of the system, exhibiting in a clear way the monotonous increase of entanglement to its final value. 

The experiment is based on linear-optics devices and a laser beam involving macroscopic photon numbers, what considerably simplifies both the state preparation and measurement setups. The results do not depend however on the beam intensity, and should remain the same in the single-photon regime.  The excellent agreement between theory and experiment evinces the suitability of this approach.

\begin{acknowledgments}
We would like to thank  Adriana Auyuanet for discussions during the development of this work. This research was supported by the Brazilian funding agencies
Funda\c c\~ao Carlos Chagas Filho de Amparo \`a Pesquisa do Estado do Rio de Janeiro (FAPERJ), Conselho Nacional de Desenvolvimento Cient\'ifico e Tecnol\'ogico (CNPq), Coordena\c c\~ao de Aperfei\c coamento de Pessoal de N\'{\i}vel Superior (CAPES), and the National Institute of Science
and Technology for Quantum Information.
\end{acknowledgments}

\bibliographystyle{apsrev4-1}
\bibliography{EnviromentInducedEntanglementExperiment}

\end{document}